\documentclass[aps, prc, floatfix, nofootinbib, superscriptaddress, twocolumn]{revtex4-1}

\usepackage{latexsym}
\usepackage{amsmath}
\usepackage{amssymb}
\usepackage{amsfonts}
\usepackage{multirow}

\usepackage{color}

\usepackage{supertabular}
\usepackage{placeins}
\usepackage{epsfig}
\usepackage{graphicx}

\definecolor{purple}{rgb}{0.5,0,0.5}
\definecolor{blue}{rgb}{0.0,0,0.9}
\definecolor{prdblue}{rgb}{0.133,0.118,0.498}
\usepackage[colorlinks=true, pdfstartview=FitV, linkcolor=prdblue, citecolor= prdblue, urlcolor=prdblue]{hyperref}

\begin{document}
\title{Analysis of hidden-bottom $bb\bar{b}\bar{b}$ states}
\author{Xiaoyun Chen}
\email{xychen@jit.edu.cn} \affiliation{Department of Basic
Courses, Jinling Institute of Technology,\\ Nanjing 211169, P. R.
China}


\begin{abstract}
Motivated by the searching for $bb\bar{b}\bar{b}$ states at LHC
recently, we calculate the ground-state energies of
$bb\bar{b}\bar{b}$ states with quantum numbers
$IJ^P=00^+,01^+,02^+$ in a nonrelativistic chiral quark model
using the Gaussian expansion method. In our calculations, two structures,
meson-meson and diquark-antidiquark, and their coupling, along with
all possible color configurations are considered. It is expected
that the studies shall be helpful for the experimental searching of
fully-heavy exotic tetraquark states.
\end{abstract}


\maketitle


\section{Introduction} \label{introduction}
In traditional quark models, meson consists of a quark and an antiquark,
and baryon is made up of three quarks. Since the first exotic resonance $X(3872)$
was announced by the Bell Collaboration in 2003~\cite{Choi:2003}, this situation has been
changed. Actually, multiquark states were proposed from the beginning of
quark model~\cite{Dyson,Jaffe1,Jaffe2,Jaffe3}.
In the past years, the charmonium-like and bottomonium-like states, the so-called $XYZ$ states
have been observed by experiments which provided us with good opportunities to study
exotic states and help us understand the strong interactions. Especially for those
charged states, they must be multiquark states consisted of two heavy quarks and two light quarks,
If they do exist. In comparison with these systems, the tetraquark states composed of four heavy
quarks, $QQ\bar{Q}\bar{Q}$, $Q=b$ quark or $c$ quark states are
much simpler because long-distance effects from light quarks are
expected not to be appreciable and the short-distance effects play
an important role now.

In this paper we will focus on the heaviest tetraquark system
involve four bottom flavored quarks, $bb\bar{b}\bar{b}$ state.
Although it is still missing in experiment, recent studies
indicate that this QCD bound state may be observable as a
resonance at the LHC in the mass range $\sim$ $18-19$ GeV in the
four lepton final state~\cite{CMS2016}. In fact, a hint has been
reported~\cite{CMS2018}. By solving the nonrelativistic
Schr\"{o}dinger equation, the mass of $bb\bar{b}\bar{b}$ state is
under the threshold of decay into a vector bottomonia pair~
\cite{prd86034004}. In Ref~\cite{yangbai2016}, Y. Bai \emph{et
al.} gave the mass of $bb\bar{b}\bar{b}$ state, which is around
100 MeV below twice the $\eta_{b}$ mass using a diffusion Monte
Carlo method. E. Eichten gave the point that such a heavy state
with a large branching fraction into $\Upsilon\Upsilon^*$ is
likely discoverable at the LHC since CMS has given the observation
of $\Upsilon$ pair production~\cite{CMS2016,Eichten2016}. M. N.
Anwar \emph{et al.} advocated the existence of $bb\bar{b}\bar{b}$
state with the predicted mass $18.72\pm0.02$ GeV~\cite{Anwar2018}.
Z. G. Wang calculated the mass of $bb\bar{b}\bar{b}$ state,
$M(0^{++})=18.84\pm0.09$ GeV and $M(2^{++})=18.85\pm0.09$ GeV with
the moment QCD sum rules~\cite{ZGwang2017}. The study by W. Chen
\emph{et al.} showed that the $bb\bar{b}\bar{b}$ states lie below
the threshold of two $\eta_b$ and they are probably stable and
very narrow~\cite{Weichen2018}.

But there also some work argues to the contrary. Hughes \emph{et
al.} found no evidence of a QCD bound tetraquark below the lowest
noninteracting thresholds using the first-principles lattice
nonrelativistic QCD methodology ~\cite{Hughes2018}. J. Wu~\emph{et
al.} estimated the masses of $bb\bar{b}\bar{b}$ states and they
are above the lowest meson-meson threshold in the framework of the
color-magnetic interaction~\cite{prd97094015}. J. -M.
Richard~\emph{et al.} also gave the results that the
$bb\bar{b}\bar{b}$ states are found to be unbound in the
constituent quark model~\cite{prd95054019}.
Ref.~\cite{prd95034011} also obtained the negative results about
these states. Recent work in Ref.~\cite{epjc782018} show that
$bb\bar{b}\bar{b}$ should be hardly visible at LHCb, given the
current sensitivity. Nevertheless, it should become observable
with higher statistics.

In this work, we try to study the ground states of the beauty-full system
with the quantum numbers $IJ^P=00^+,01^+,02^+$ in a
nonrelativistic chiral quark model with the help of Gaussian
expansion method (GEM)~\cite{Hiyama:2003cu}. The pure meson-meson
and pure diquark-antidiquark structure, and the coupling of two
structures are considered respectively, along with all possible
color configurations. For the interaction between the heavy
quarks, the short-range gluon exchange force is a dominant source.
Besides, we discuss the possible employment of the effective meson
exchange between heavy quarks in the context of chiral quark model and
try to look for attractive mechanism in the system.

The paper is organized as follows. After the introduction, we will
simply introduce the chiral quark model and how to construct the
wave functions of four-quark states. In Sec.~\ref{Numerical
Results}, our numerical results and discussion are presented. In
Sec.~\ref{epilogue}, a brief summary is given.


\section{Chiral quark model and wave functions of four-quark states}
\label{GEM and chiral quark model}
\subsection{The chiral quark model}
The chiral quark model has been successful both in describing the
hadron spectra and hadron-hadron interactions. The details of the
model can be found in Ref.~\cite{094016chen,Vijande:2005}. For
$b\bar{b}b\bar{b}$ full-heavy system, the Hamiltonian of the
chiral quark model consists of three parts: quark rest mass,
kinetic energy, and potential energy:
\begin{align}
 H & = \sum_{i=1}^4 m_i  +\frac{p_{12}^2}{2\mu_{12}}+\frac{p_{34}^2}{2\mu_{34}}
  +\frac{p_{1234}^2}{2\mu_{1234}}  \quad  \nonumber \\
  & + \sum_{i<j=1}^4 \left( V_{ij}^{C}+V_{ij}^{G}\right).
\end{align}
The potential energy consists of pieces describing quark
confinement (C); one-gluon-exchange (G). The forms of potentials
are shown below (only central parts are presented)
\cite{094016chen}: {\allowdisplaybreaks
\begin{subequations}
\begin{align}
V_{ij}^{C}&= ( -a_c r_{ij}^2-\Delta ) \boldsymbol{\lambda}_i^c
\cdot \boldsymbol{\lambda}_j^c ,  \\
 V_{ij}^{G}&= \frac{\alpha_s}{4} \boldsymbol{\lambda}_i^c \cdot \boldsymbol{\lambda}_{j}^c
\left[\frac{1}{r_{ij}}-\frac{2\pi}{3m_im_j}\boldsymbol{\sigma}_i\cdot
\boldsymbol{\sigma}_j
  \delta(\boldsymbol{r}_{ij})\right],  \\
\delta{(\boldsymbol{r}_{ij})} & =
\frac{e^{-r_{ij}/r_0(\mu_{ij})}}{4\pi r_{ij}r_0^2(\mu_{ij})}.
%
\end{align}
\end{subequations}}
\hspace*{-0.5\parindent}%
%
$m$ is the constituent masse of quark/antiquark, and $\mu_{ij}$ is the reduced masse
of two interacting quarks and
\begin{equation}
\mu_{1234}=\frac{(m_1+m_2)(m_3+m_4)}{m_1+m_2+m_3+m_4};
\end{equation}
$\mathbf{p}_{ij}=(\mathbf{p}_i-\mathbf{p}_j)/2$,
$\mathbf{p}_{1234}= (\mathbf{p}_{12}-\mathbf{p}_{34})/2$;
$r_0(\mu_{ij}) =s_0/\mu_{ij}$; $\boldsymbol{\sigma}$ are the
$SU(2)$ Pauli matrices; $\boldsymbol{\lambda}$,
$\boldsymbol{\lambda}^c$ are $SU(3)$ flavor, color Gell-Mann
matrices, respectively;
%
%
and $\alpha_s$ is an effective scale-dependent running coupling
\cite{Vijande:2005},
\begin{equation}
\alpha_s(\mu_{ij})=\frac{\alpha_0}{\ln\left[(\mu_{ij}^2+\mu_0^2)/\Lambda_0^2\right]}.
\end{equation}
All the parameters are determined by fitting the meson spectrum,
from light to heavy;
and the resulting values are listed in
Table~\ref{modelparameters}.

\begin{table}[!t]
\begin{center}
\caption{ \label{modelparameters} Model parameters, determined by
fitting the meson spectrum from light to heavy.}
\begin{tabular}{llr}
\hline\noalign{\smallskip}
Quark masses   &$m_u=m_d$    &313  \\
   (MeV)       &$m_s$         &536  \\
               &$m_c$         &1728 \\
               &$m_b$         &5112 \\
\hline
Confinement        &$a_c$ (MeV fm$^{-2}$)         &101 \\
                   &$\Delta$ (MeV)     &-78.3 \\
\hline
OGE                 & $\alpha_0$        &3.67 \\
                   &$\Lambda_0({\rm fm}^{-1})$ &0.033 \\
                  &$\mu_0$(MeV)    &36.98 \\
                   &$s_0$(MeV)    &28.17 \\
\hline
\end{tabular}
\end{center}
\end{table}

\subsection{The wave functions of four-quark states}
The wave functions of four-quark states for the two structures,
diquark-antidiquark and meson-meson, can be constructed in two steps. For each degree
of freedom, first we construct the wave functions for two-body
clusters, then coupling the wave functions of two clusters to the
wave functions of four-quark states.

(1) Diqaurk-antidiquark structure.

For spin part, the wave functions for two-body clusters are,
\begin{align}
&\chi_{11}=\alpha\alpha,~~
\chi_{10}=\frac{1}{\sqrt{2}}(\alpha\beta+\beta\alpha),~~
\chi_{1-1}=\beta\beta,\nonumber \\
&\chi_{00}=\frac{1}{\sqrt{2}}(\alpha\beta-\beta\alpha),
\end{align}
then the wave functions for four-quark states are obtained,
 {\allowdisplaybreaks
\begin{subequations}\label{spinwavefunctions}
\begin{align}
\chi_{0}^{\sigma
1}&=\chi_{00}\chi_{00},\\
\chi_{0}^{\sigma
2}&=\sqrt{\frac{1}{3}}(\chi_{11}\chi_{1-1}-\chi_{10}\chi_{10}+\chi_{1-1}\chi_{11}),\\
\chi_{1}^{\sigma
3}&=\chi_{00}\chi_{11},\\
 \chi_{1}^{\sigma
4}&=\chi_{11}\chi_{00},\\
\chi_{1}^{\sigma
5}&=\frac{1}{\sqrt{2}}(\chi_{11}\chi_{10}-\chi_{10}\chi_{11}),\\
\chi_{2}^{\sigma 6}&=\chi_{11}\chi_{11},
\end{align}
\end{subequations}}
where the subscript of $\chi$ represents the total spin of four-quark states, it takes the values
$S=0, 1, 2$, and only one component is shown for a given total spin $S$.

For flavor part, the wave function for $bb\bar{b}\bar{b}$ state is very simple
\begin{equation}
\chi_{d0}^{f} =bb\bar{b}\bar{b},
\end{equation}
the subscript of $\chi$ represents the isospin of $bb\bar{b}\bar{b}$, and it takes $I=0$.

For color part, the wave functions of four-quark states must be color singlet $[222]$ and
it is obtained as below,
\begin{subequations}
\begin{align}
\chi^{c}_{d1} & =
\frac{\sqrt{3}}{6}(rg\bar{r}\bar{g}-rg\bar{g}\bar{r}+gr\bar{g}\bar{r}-gr\bar{r}\bar{g} \nonumber \\
&~~~+rb\bar{r}\bar{b}-rb\bar{b}\bar{r}+br\bar{b}\bar{r}-br\bar{r}\bar{b} \nonumber \\
&~~~+gb\bar{g}\bar{b}-gb\bar{b}\bar{g}+bg\bar{b}\bar{g}-bg\bar{g}\bar{b}).  \\
\chi^{c}_{d2}&=\frac{\sqrt{6}}{12}(2rr\bar{r}\bar{r}+2gg\bar{g}\bar{g}+2bb\bar{b}\bar{b}
    +rg\bar{r}\bar{g}+rg\bar{g}\bar{r} \nonumber \\
&~~~+gr\bar{g}\bar{r}+gr\bar{r}\bar{g}+rb\bar{r}\bar{b}+rb\bar{b}\bar{r}+br\bar{b}\bar{r} \nonumber \\
&~~~+br\bar{r}\bar{b}+gb\bar{g}\bar{b}+gb\bar{b}\bar{g}+bg\bar{b}\bar{g}+bg\bar{g}\bar{b}).
\end{align}
\end{subequations}
Where, $\chi_{d1}^{c}$ and $\chi_{d2}^{c}$ represents the color antitriplet-triplet ($\bar{3}\times3$)
and sextet-antisextet ($6\times\bar{6}$) coupling, respectively. The detailed coupling process
for the color wave functions can refer to our previous work \cite{054022chen}.

(2) Meson-meson structure.

For spin part, the wave functions are the same as those of the diquark-antidiquark structure,
Eq.~(\ref{spinwavefunctions}).

The flavor wave function of $bb\bar{b}\bar{b}$ state takes as follows,
\begin{equation}
\chi_{m0}^{f} =\bar{b}b\bar{b}b,
\end{equation}
the subscript of $\chi_{m0}$ represents the isospin of four-quark states, $I=0$.

For color part, the wave functions of four-quark states in the meson-meson structure are,
\begin{subequations}
\begin{align}
\chi_{m1}^{c}&=\frac{1}{3}(\bar{r}r+\bar{g}g+\bar{b}b)(\bar{r}r+\bar{g}g+\bar{b}b),\\
\chi_{m2}^{c}&=\frac{\sqrt{2}}{12}(3\bar{b}r\bar{r}b+3\bar{g}r\bar{r}g+3\bar{b}g\bar{g}b+3\bar{g}b\bar{b}g+3\bar{r}g\bar{g}r \nonumber \\
&~~~+3\bar{r}b\bar{b}r+2\bar{r}r\bar{r}r+2\bar{g}g\bar{g}g+2\bar{b}b\bar{b}b-\bar{r}r\bar{g}g \nonumber\\
&~~~-\bar{g}g\bar{r}r-\bar{b}b\bar{g}g-\bar{b}b\bar{r}r-\bar{g}g\bar{b}b-\bar{r}r\bar{b}b).
\end{align}
\end{subequations}
Where, $\chi_{m1}^{c}$ and $\chi_{m2}^{c}$ represents the color singlet-singlet($1\times1$) and
color octet-octet($8\times8$) coupling, respectively. The details refer to our previous work \cite{054022chen}.

As for the orbital wave functions, they can be constructed by coupling the orbital wave function
for each relative motion of the system,
\begin{align}\label{spatialwavefunctions}
\Psi_{L}^{M_{L}}=\left[[\Psi_{l_1}({\bf r}_{12})\Psi_{l_2}({\bf
r}_{34})]_{l_{12}}\Psi_{L_r}({\bf r}_{1234}) \right]_{L}^{M_{L}},
\end{align}
where $l_1$ and $l_2$ is the angular momentum of two clusters,
respectively. $\Psi_{L_r}(\mathbf{r}_{1234})$ is the wave function
of the relative motion between two sub-clusters with orbital
angular momentum $L_r$. $L$ is the total orbital angular momentum
of four-quark states. Here for the low-lying $bb\bar{b}\bar{b}$
state, all angular momentum ($l_1, l_2, L_r, L$) are taken as
zero. The Jacobi coordinates are defined as,
\begin{align}\label{jacobi}
{\bf r}_{12}&={\bf r}_1-{\bf r}_2, \nonumber \\
{\bf r}_{34}&={\bf r}_3-{\bf r}_4, \nonumber\\
{\bf r}_{1234}&=\frac{m_1{\bf r}_1+m_2{\bf
r}_2}{m_1+m_2}-\frac{m_3{\bf r}_3+m_4{\bf r}_4}{m_3+m_4}.
\end{align}
For diquark-antidiquark structure, the quarks are numbered as $1, 2$, and the antiquarks are
numbered as $3, 4$; for meson-meson structure, the quark and antiquark in one cluster are marked as
$1, 2$, the other quark and antiquark are marked as $3, 4$. In the two structure coupling calculation,
the indices of quarks, antiquarks in diquark-antidiquark structure will be changed to be consistent with
the numbering scheme in meson-meson structure.
In GEM, the spatial wave function is expanded by Gaussians~\cite{Hiyama:2003cu}:
\begin{subequations}
\label{radialpart}
\begin{align}
\Psi_{l}^{m}(\mathbf{r}) & = \sum_{n=1}^{n_{\rm max}} c_{n}\psi^G_{nlm}(\mathbf{r}),\\
\psi^G_{nlm}(\mathbf{r}) & = N_{nl}r^{l}
e^{-\nu_{n}r^2}Y_{lm}(\hat{\mathbf{r}}),
\end{align}
\end{subequations}
where $N_{nl}$ are normalization constants,
\begin{align}
N_{nl}=\left[\frac{2^{l+2}(2\nu_{n})^{l+\frac{3}{2}}}{\sqrt{\pi}(2l+1)}
\right]^\frac{1}{2}.
\end{align}
$c_n$ are the variational parameters, which are determined
dynamically. The Gaussian size parameters are chosen according to
the following geometric progression
\begin{equation}\label{gaussiansize}
\nu_{n}=\frac{1}{r^2_n}, \quad r_n=r_1a^{n-1}, \quad
a=\left(\frac{r_{n_{\rm max}}}{r_1}\right)^{\frac{1}{n_{\rm
max}-1}}.
\end{equation}
This procedure enables optimization of the ranges using just a
small number of Gaussians. Finally, the complete channel wave
function for the four-quark system for diquark-antidiquark
structure is written as
\begin{align}\label{diquarkpsi}
&\Psi_{IJ,i,j}^{M_IM_J}={\cal A}_1[\Psi_{L}\chi_S^{\sigma i}]_{J}^{M_J}\chi_{d0}^{f}\chi^{c}_{dj},\nonumber \\
&(i=1\sim6; j=1,2; S=0,1,2),
\end{align}
where ${\cal A}_1$ is the antisymmetrization operator,
\begin{equation}
{\cal A}_1=\frac{1}{2}(1-P_{12}-P_{34}+P_{12}P_{34}).
\end{equation}

For meson-meson structure, the complete wave function is written
as
\begin{align}
&\Psi_{IJ,i,j}^{M_IM_J}= {\cal A}_2[\Psi_{L}\chi_S^{\sigma
i}]_{J}^{M_J}\chi_{m0}^{f}\chi^{c}_{mj},\nonumber \label{mesonpsi}\\
&(i=1\sim6; j=1,2; S=0,1,2),
\end{align}
where ${\cal A}_2$ is the antisymmetrization operator,
\begin{equation}
{\cal A}_2=\frac{1}{2}(1-P_{13}-P_{24}+P_{13}P_{24}).
\end{equation}

Lastly, the eigenenergies of the four-quark system are obtained by
solving a Schr\"{o}dinger equation:
\begin{equation}
    H \, \Psi^{\,M_IM_J}_{IJ}=E^{IJ} \Psi^{\,M_IM_J}_{IJ},
\end{equation}
where $\Psi^{\,M_IM_J}_{IJ}$ is the wave function of the
four-quark states, which is the linear combinations of the above
channel wave functions, Eq.~(\ref{diquarkpsi}) in the
diquark-antidiquark structure or Eq.~(\ref{mesonpsi}) in the
meson-meson structure, respectively.

\section{Numerical Results and discussions}
\label{Numerical Results}

In the framework of the chiral quark model, we calculated the
masses of the four-quark state $bb\bar{b}\bar{b}$ with quantum
numbers $IJ^P=00^+,01^+,02^+$. The meson-meson and
diquark-antidiqaurk structure, and the coupling of these two
structures are first considered. For each structure, all possible
color configurations and their coupling are taken into account.
i.e., for meson-meson structure ($\bar{b}b\bar{b}b$), two color
configurations, color singlet-singlet ($1\times1$), octet-octet
($8\times8$) and the mixture of them are studied. For
diquark-antidiquark structure ($bb\bar{b}\bar{b}$), also two color
configurations, antitriplet-triplet ($\bar{3}\times3$),
sextet-antisextet ($6\times\bar{6}$) and their coupling are
considered.
\begin{figure}[t]
\centerline{\includegraphics[scale=0.2]{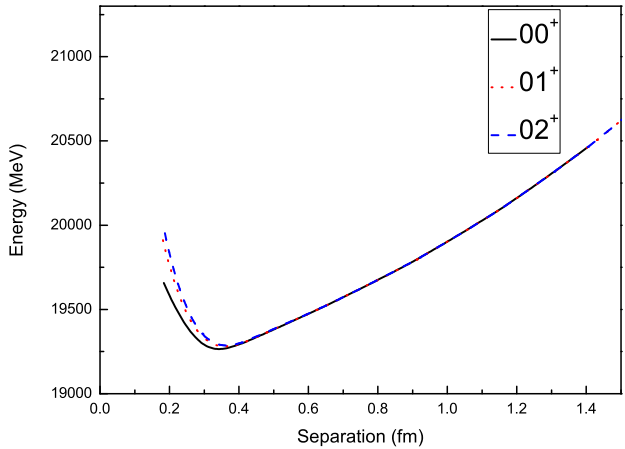}}
\caption{\label{veff} The lowest eigeneneries of $00^+$,
$01^+$ and $02^+$ state as a function of the distance between the
diquark and antidiquark in adiabatic approximation.}
\end{figure}

\subsection{No Goldstone Boson Exchanges}

For beauty-full system, generally the Goldstone boson exchanges
between $b$ quarks are not introduced because of the large mass of
$b$-quark. The eigenvalues of $bb\bar{b}\bar{b}$ four-quark state
in meson-meson and diquark-antidiqaurk structures are demonstrated
in Table~\ref{nomesonexchanges1} and
Table~\ref{nomesonexchanges2}, respectively. In
Table~\ref{nomesonexchanges1}, the column with head "channel"
represents the index of the antisymmetrized wave functions of
$bb\bar{b}\bar{b}$ state. $E_0$ is the eigenenergy of each
channel. $E_{cc1}$ gives the eigenenergy with the channel coupling
of the two color configurations ($1\times1$ and $8\times8$), and
from the table, we can see that the effect of the hidden color
channel is too tiny to be visible. Coupling all different
spin-color configurations, we get the eigenvalue ($E_{cc2}$) for
each set of quantum numbers ($00^+$,$01^+$,$02^+$). The results
indicate that the couplings effect are also very small. All the
eigenvalues are higher than the theoretical thresholds. With
increasing range, all the eigenvalues are approaching to the
theoretical thresholds. So we found no bound states for
$bb\bar{b}\bar{b}$ state in meson-meson structure.

For diquark-antidiquark structure, in
Table~\ref{nomesonexchanges2}, we can see that the couplings of
the two color configurations ($\bar{3}\times3$ and
$6\times\bar{6}$) are rather strong. But the eigenvalue of each
state is still higher than the corresponding theoretical
threshold. Because the colorful clusters cannot fall apart, there
may be a resonance even with the higher eigenenergy. To check this
possibility, we preform an adiabatic calculations for the $00^+$,
$01^+$ and $02^+$ states. In this case, the number of the
Gaussians used for the relative motion between the diquark and
antidiquark subclusters is limited to 1. The lowest adiabatic
eigenenergies of these states with different separation between
two subclusters are shown in Fig~\ref{veff}. It reveals that the
energies are increasing when the separation increases. And at the
separation about $0.3$ fm, there comes the minimum energies for
these states which manifests that the subclusters are not willing
to be too close or falling apart. And in our calculations, in pure
diquark-antidiquark structure the $bb\bar{b}\bar{b}$ tetraquark
state may be a resonance state with the lowest mass 19177.5 MeV
which is a little larger than the bound state values 18827 MeV
~\cite{prd97094015} and $18826\pm25$ MeV~\cite{prd95034011}.

When considering the coupling of two structures, the results are
shown in Table ~\ref{nomesonexchanges-mix}. $E_1$ represents the
low-lying eigenvalues. For each state, we found that the low-lying
energy tends to be the same with those in the pure meson-meson
structure, which indicates that there is still no bound state
after considering the mixtures of two structures. In order to
looking for the possible resonances, we calculated the distance
between $b$ and $\bar{b}$ quark, denotes as $R_{b\bar{b}}$ in
Table ~\ref{nomesonexchanges-mix}, as well as the distance between
$b$ and $b$ quark, denotes as $R_{bb}$ for each eigenstate.
$E_{2}$ represents the eigenenergy of the first possible resonance
state. From the table, we can find that the lowest mass of the
possible resonance state is reduced to 18872.8 MeV, which is close
to the previous results~\cite{prd97094015,prd95034011}. Compared
with 19177.5 MeV in pure diquark-antidiquark structure, we can see
that the coupling of the two structures plays an important role.
The percentage of each structure is not meaningful because there
is a large overlap between two structures~\cite{Ji}. For $01^+$
and $02^+$ state, we cannot find the relative stable resonance
states around 19 GeV.

\begin{table}[!t]
\begin{center}
\caption{ \label{nomesonexchanges1} The eigenenergies of
$bb\bar{b}\bar{b}$ state for meson-meson structure with no
Goldstone boson exchanges (unit: MeV). $E_{th1}$ and $E_{th2}$
represents the theoretical and experimental threshold of each
channel, respectively.}
\begin{tabular}{ccccccc} \hline \hline
$IJ^P$~~~&channel&~~~$E_0$~~~&$E_{cc1}$~~~&$E_{cc2}$~~~&$E_{th1}$~~~&$E_{th2}$ \\
\hline
$00^+$ &$\chi^{\sigma 1}_{0}\chi^{f}_{m0}\chi^{c}_{m1}$  &18669.6  &18669.6 &18669.6  &18669.3  &18798.0 \\
       &$\chi^{\sigma 1}_{0}\chi^{f}_{m0}\chi^{c}_{m2}$  &19205.4  &        &         &         & \\
       &$\chi^{\sigma 2}_{0}\chi^{f}_{m0}\chi^{c}_{m1}$  &18928.3  &18928.3 &         &18927.8  &18920.6 \\
       &$\chi^{\sigma 2}_{0}\chi^{f}_{m0}\chi^{c}_{m2}$  &19194.9  &        &         &         &   \\
       \hline
$01^+$ &$\chi^{\sigma 3}_{1}\chi^{f}_{m0}\chi^{c}_{m1}$  &18798.9 &18798.9 &18798.9   &18798.6  &18859.3 \\
       &$\chi^{\sigma 3}_{1}\chi^{f}_{m0}\chi^{c}_{m2}$  &19179.2  &        &         &         & \\
       &$\chi^{\sigma 4}_{1}\chi^{f}_{m0}\chi^{c}_{m1}$  &18798.9 &18798.9  &         &18798.6  &18859.3 \\
       &$\chi^{\sigma 4}_{1}\chi^{f}_{m0}\chi^{c}_{m2}$  &19179.2  &        &         &         & \\
       \hline
$02^+$ &$\chi^{\sigma 6}_{2}\chi^{f}_{m0}\chi^{c}_{m1}$  &18928.3  &18928.3 &18928.3  &18927.8  &18920.6\\
       &$\chi^{\sigma 6}_{2}\chi^{f}_{m0}\chi^{c}_{m2}$  &19195.0  &        &         &         & \\
\hline \hline
\end{tabular}
%
\caption{ \label{nomesonexchanges2} The eigenenergies of
$bb\bar{b}\bar{b}$ state for diquark-antidiquark structure with no
Goldstone boson exchanges (unit: MeV). $E_{th1}$ represents the
theoretical threshold of each channel.}
\begin{tabular}{ccccc} \hline \hline
$IJ^P$~~~&channel&~~~$E_0$~~~&$E_{cc1}$~~~&$E_{th1}$~~~ \\
\hline
$00^+$ &$\chi^{\sigma 1}_{0}\chi^{f}_{d0}\chi^{c}_{d2}$  &19191.1  &19177.5 &18669.3  \\
       &$\chi^{\sigma 2}_{0}\chi^{f}_{d0}\chi^{c}_{d1}$  &19221.0  &        &         \\
       \hline
$01^+$ &$\chi^{\sigma 5}_{1}\chi^{f}_{d0}\chi^{c}_{d1}$  &19226.8  &19226.8 &18927.8  \\
       \hline
$02^+$ &$\chi^{\sigma 6}_{2}\chi^{f}_{d0}\chi^{c}_{d1}$  &19237.6  &19237.6 &18927.8  \\
\hline \hline
\end{tabular}

\caption{ \label{nomesonexchanges-mix} The eigenenergies of
$bb\bar{b}\bar{b}$ state after considering the coupling of two
structures with no Goldstone boson exchanges.}
\begin{tabular}{ccccc} \hline \hline
$IJ^P$~~~~~&$E_1$(MeV)~~~~~&$E_2$(MeV)~~~~~&$R_{b\bar{b}}$(fm)~~~~~&$R_{bb}$(fm) \\
\hline
$00^+$ &18669.6  &18872.8   &0.58  &0.85    \\
$01^+$ &18798.9  &...       &...  &...    \\
$02^+$ &18928.3  &...       &...  &...    \\
\hline \hline
\end{tabular}
\end{center}
\end{table}

\begin{table}[!t]
\begin{center}
\caption{ \label{etab1} The eigenenergies of $bb\bar{b}\bar{b}$
state for meson-meson structure only with heavy meson $\eta_{b}$
exchange (unit: MeV). $E_{th1}$ and $E_{th2}$ represents the
theoretical and experimental threshold of each channel,
respectively.}
\begin{tabular}{ccccc} \hline \hline
$IJ^P$ & channel & $E_{cc1}$ & $E_{th1}$ & $E_{th2}$ \\
\hline
$00^+$  &$\chi^{\sigma 1}_{0}\chi^{f}_{m0}\chi^{c}_{m1}$, $\chi^{\sigma 1}_{0}\chi^{f}_{m0}\chi^{c}_{m2}$  &18704.3    &18704.0  &18798.0 \\
        &$\chi^{\sigma 2}_{0}\chi^{f}_{m0}\chi^{c}_{m1}$, $\chi^{\sigma 2}_{0}\chi^{f}_{m0}\chi^{c}_{m2}$  &18924.1    &18923.8  &18920.6 \\
$01^+$  &$\chi^{\sigma 3}_{1}\chi^{f}_{m0}\chi^{c}_{m1}$, $\chi^{\sigma 3}_{1}\chi^{f}_{m0}\chi^{c}_{m2}$  &18814.2    &18813.9  &18859.3 \\
        &$\chi^{\sigma 4}_{1}\chi^{f}_{m0}\chi^{c}_{m1}$, $\chi^{\sigma 4}_{1}\chi^{f}_{m0}\chi^{c}_{m2}$  &18814.2    &18813.9  &18859.3 \\
$02^+$  &$\chi^{\sigma 6}_{2}\chi^{f}_{m0}\chi^{c}_{m1}$, $\chi^{\sigma 6}_{2}\chi^{f}_{m0}\chi^{c}_{m2}$  &18924.1    &18923.8  &18920.6 \\
\hline \hline
\end{tabular}
\end{center}
\end{table}

\begin{table}[!t]
\begin{center}
\caption{ \label{etab2} The eigenenergies of $bb\bar{b}\bar{b}$
state for diquark-antidiquark structure only with heavy meson
$\eta_{b}$ exchange (unit: MeV). $E_{th1}$ and $E_{th2}$
represents the theoretical and experimental threshold of each
channel, respectively.}
\begin{tabular}{ccccc} \hline \hline
$IJ^P$ & channel & $E_{cc1}$ & $E_{th1}$ & $E_{th2}$ \\
\hline
$00^+$ &$\chi^{\sigma 1}_{0}\chi^{f}_{d0}\chi^{c}_{d2}$, $\chi^{\sigma 2}_{0}\chi^{f}_{d0}\chi^{c}_{d1}$ &19177.9   &18704.0  &18798.0 \\
$01^+$ &$\chi^{\sigma 5}_{1}\chi^{f}_{d0}\chi^{c}_{d1}$ &19226.4   &18923.8  &18920.6 \\
$02^+$ &$\chi^{\sigma 6}_{2}\chi^{f}_{d0}\chi^{c}_{d1}$ &19235.5   &18923.8  &18920.6 \\
\hline \hline
\end{tabular}
\end{center}
\end{table}

\subsection{Inclusion of $\eta_b$ Exchange}

If the bound $bb\bar{b}\bar{b}$ state is found by experiments, quark model has to be
expanded to account for the challenge. On hadron level, to study mutliquark systems,
the heavy meson exchanges are invoked. In studying hidden-charm pentaquark states $N^{*}$
and $\Lambda^{*}$, the SU(4) flavor symmetry was employed~\cite{Wu}. The exchange of
heavy vector mesons was also used in exploring the charmonium-like hadrons~\cite{Aceti,He}.
Here we extend the chiral quark model by including the heavy meson exchange between
$b(\bar{b})$ quarks, to check whether the heavy meson exchange can provide attractive mechanism.
First, the $\eta_b$ exchange is introduced,
\begin{eqnarray}
V_{ij}^{\eta_b} & = & \frac{g_{ch}^2}{4\pi}\frac{m_{\eta_b}^2}{12m_im_j}
  \frac{\Lambda_{\eta_b}^2}{\Lambda_{\eta_b}^2-m_{\eta_b}^2}m_{\eta_b} \nonumber  \\
 & & \left[ Y(m_\chi r_{ij})-\frac{\Lambda_{\chi}^3}{m_{\chi}^3}Y(\Lambda_{\chi} r_{ij})
  \right] \lambda_i^{15} \lambda_j^{15} \boldsymbol{\sigma}_i \cdot\boldsymbol{\sigma}_j,
\end{eqnarray}
where $\lambda_j^{15}=diag(1,1,1,1,-4)/\sqrt{10}$ and
$Y(x)=e^{-x}/x$. The results with the inclusion of the $\eta_b$
exchange
are shown in Tables~\ref{etab1} and \ref{etab2}. As aforementioned, the coupling effects of the
different spin-color configurations are very small, the inclusion of $\eta_b$ exchange does not change
the statement. So in Tables~\ref{etab1} and \ref{etab2}, we just give the eigenvalues $E_{cc1}$ for
conciseness. In meson-meson structure (Table~\ref{etab1}), we can see that the effect of $\eta_b$ exchange
is too small due to its large mass and
the eigenvalue for each state is still higher than and approaching to the corresponding
theoretical threshold. No bound states are found.

For diquark-antidiquark structure (Table~\ref{etab2}), the eigenvalues for these states
are similar with those without $\eta_b$ exchange, a resonance may be exist, rather than
a meson-meson molecular state.


\subsection{Invoking of an effective $\sigma$ exchange}
The chiral partner, scalar $\sigma$ meson provide a universal attraction in the $u,d$ systems.
Extended to $u,d,s$ three-flavor world, the scalar nonet states are used. For the heavy quark
systems, it is a possible way to introduce the scalar meson exchange to increase the attraction
between quarks. Clearly to find $n^2$ scalar mesons, which associated with SU(n) flavor symmetry,
is impractical. Instead often an effective $\sigma$ meson exchange is used~\cite{sigma1,sigma2}.
Now we calculated the eigenenergies of the $bb\bar{b}\bar{b}$ state by considering an effective
$\sigma$ meson exchange with the mass of $\sigma$ taking a series of values, 1.0 GeV, 1.5 GeV,
2.5 GeV, 3.5 GeV. The results for meson-meson and diquark-antidiquark structures are demonstrated
in Tables~\ref{sigma1} and \ref{sigma2}, respectively. In the tables, the $E_B$ gives the binding
energy of the state. For meson-meson structure, from Table~\ref{sigma1}, the eigenenergies of the
states are all lower than the corresponding thresholds. The binding energies are getting smaller
with the increasing of the $\sigma$ mass.

For diquark-antidiquark structure, the results are shown in
table~\ref{sigma2}, where ``..." means that the eigenvalue is
higher than the corresponding threshold, which indicates there's
no bound state. From the table, we can find that for $00^+$ state,
no matter what the mass of $\sigma$ is, no bound states are found.
For $01^+$ and $02^+$ state, only when the mass of $\sigma$ takes
the small values, 1.0 GeV, 1.5 GeV, we can find bound states.

From the above results, we can see that extra attractive potential
must be introduced if bound $bb\bar{b}\bar{b}$ states exist. Introducing an
effective $\sigma$ meson exchange is a common and economic way to
increase the attraction between clusters. However, too few scalar
are found experimentally for the heavy flavor systems, and the use of
effective $\sigma$ meson exchange is too artificial.

\begin{table}[!t]
\begin{center}
\caption{ \label{sigma1} The eigenenergies of $bb\bar{b}\bar{b}$
state for meson-meson structure only with an effective $\sigma$
exchange (unit: MeV).}
\begin{tabular}{ccccccccccc} \hline \hline
$IJ^P$ ~~~~&channel~~~~&$E_{cc1}$~~~~&$E_{th_1}$~~~~ &~~~$E_B$~~~~ \\
\hline
&  &  \multicolumn{3}{c}{$m_{\sigma}$=1000} \\ \cline{3-5}

$00^+$ &$\chi^{\sigma 1}_{0}\chi^{f}_{m0}\chi^{c}_{m1}$,$\chi^{\sigma 1}_{0}\chi^{f}_{m0}\chi^{c}_{m2}$   &18107.5 &18337.8  &$-230.3$ \\ &$\chi^{\sigma 2}_{0}\chi^{f}_{m0}\chi^{c}_{m1}$,$\chi^{\sigma 2}_{0}\chi^{f}_{m0}\chi^{c}_{m2}$   &18381.8 &18650.4 &$-268.6$ \\
$01^+$ &$\chi^{\sigma 3}_{1}\chi^{f}_{m0}\chi^{c}_{m1}$,$\chi^{\sigma 3}_{1}\chi^{f}_{m0}\chi^{c}_{m2}$   &18288.2 &18494.1 &$-205.9$ \\
       &$\chi^{\sigma 4}_{1}\chi^{f}_{m0}\chi^{c}_{m1}$,$\chi^{\sigma 4}_{1}\chi^{f}_{m0}\chi^{c}_{m2}$   &18288.2 &18494.1 &$-205.9$ \\
$02^+$ &$\chi^{\sigma 6}_{2}\chi^{f}_{m0}\chi^{c}_{m1}$,$\chi^{\sigma 6}_{2}\chi^{f}_{m0}\chi^{c}_{m2}$   &18425.4 &18650.4 &$-225.0$  \\
\hline
 &  &  \multicolumn{3}{c}{$m_{\sigma}$=1500} \\ \cline{3-5}
$00^+$ &$\chi^{\sigma 1}_{0}\chi^{f}_{m0}\chi^{c}_{m1}$,$\chi^{\sigma 1}_{0}\chi^{f}_{m0}\chi^{c}_{m2}$  &18136.9 &18321.0 &$-184.1$ \\
&$\chi^{\sigma 2}_{0}\chi^{f}_{m0}\chi^{c}_{m1}$,$\chi^{\sigma 2}_{0}\chi^{f}_{m0}\chi^{c}_{m2}$  &18429.6 &18654.0 &$-224.4$ \\
$01^+$ &$\chi^{\sigma 3}_{1}\chi^{f}_{m0}\chi^{c}_{m1}$,$\chi^{\sigma 3}_{1}\chi^{f}_{m0}\chi^{c}_{m2}$  &18330.9 &18487.5 &$-156.6$ \\
       &$\chi^{\sigma 4}_{1}\chi^{f}_{m0}\chi^{c}_{m1}$,$\chi^{\sigma 4}_{1}\chi^{f}_{m0}\chi^{c}_{m2}$  &18330.9 &18487.5 &$-156.6$ \\
$02^+$ &$\chi^{\sigma 6}_{2}\chi^{f}_{m0}\chi^{c}_{m1}$,$\chi^{\sigma 6}_{2}\chi^{f}_{m0}\chi^{c}_{m2}$ &18475.2 &18654.0 &$-178.8$ \\
\hline
 &  &  \multicolumn{3}{c}{$m_{\sigma}$=2500} \\ \cline{3-5}
$00^+$ &$\chi^{\sigma 1}_{0}\chi^{f}_{m0}\chi^{c}_{m1}$,$\chi^{\sigma 1}_{0}\chi^{f}_{m0}\chi^{c}_{m2}$   &18255.1 &18335.2 &$-80.1$ \\
       &$\chi^{\sigma 2}_{0}\chi^{f}_{m0}\chi^{c}_{m1}$,$\chi^{\sigma 2}_{0}\chi^{f}_{m0}\chi^{c}_{m2}$   &18571.7 &18692.8 &$-121.1$ \\
$01^+$ &$\chi^{\sigma 3}_{1}\chi^{f}_{m0}\chi^{c}_{m1}$,$\chi^{\sigma 3}_{1}\chi^{f}_{m0}\chi^{c}_{m2}$   &18461.1 &18514.0 &$-52.9$ \\
       &$\chi^{\sigma 4}_{1}\chi^{f}_{m0}\chi^{c}_{m1}$,$\chi^{\sigma 4}_{1}\chi^{f}_{m0}\chi^{c}_{m2}$   &18461.1 &18514.0 &$-52.9$ \\
$02^+$ &$\chi^{\sigma 6}_{2}\chi^{f}_{m0}\chi^{c}_{m1}$,$\chi^{\sigma 6}_{2}\chi^{f}_{m0}\chi^{c}_{m2}$  &18619.1 &18692.8 &$-73.7$ \\
\hline
 &  &  \multicolumn{3}{c}{$m_{\sigma}$=3500} \\ \cline{3-5}
$00^+$ &$\chi^{\sigma 1}_{0}\chi^{f}_{m0}\chi^{c}_{m1}$,$\chi^{\sigma 1}_{0}\chi^{f}_{m0}\chi^{c}_{m2}$  &18358.5 &18371.8 &$-13.3$\\
       &$\chi^{\sigma 2}_{0}\chi^{f}_{m0}\chi^{c}_{m1}$,$\chi^{\sigma 2}_{0}\chi^{f}_{m0}\chi^{c}_{m2}$  &18688.8 &18736.0 &$-47.2$\\
$01^+$ &$\chi^{\sigma 3}_{1}\chi^{f}_{m0}\chi^{c}_{m1}$,$\chi^{\sigma 3}_{1}\chi^{f}_{m0}\chi^{c}_{m2}$   &18553.3 &18553.9 &$-0.6$\\
       &$\chi^{\sigma 4}_{1}\chi^{f}_{m0}\chi^{c}_{m1}$,$\chi^{\sigma 4}_{1}\chi^{f}_{m0}\chi^{c}_{m2}$   &18553.3 &18553.9 &$-0.6$ \\
$02^+$ &$\chi^{\sigma 6}_{2}\chi^{f}_{m0}\chi^{c}_{m1}$,$\chi^{\sigma 6}_{2}\chi^{f}_{m0}\chi^{c}_{m2}$  &18728.4 &18736.0 &$-7.6$ \\
\hline \hline
\end{tabular}
\end{center}
\end{table}

\begin{table}[!t]
\begin{center}
\caption{ \label{sigma2} The eigenenergies of $bb\bar{b}\bar{b}$
state for diquark-antidiquark structure only with an effective
$\sigma$ exchange (unit: MeV).}
\begin{tabular}{ccccccccccc} \hline \hline
$IJ^P$ ~~~~&channel~~~~&$E_{cc1}$~~~~&$E_{th_1}$~~~~ &~~~$E_B$ \\
\hline
 &  &  \multicolumn{3}{c}{$m_{\sigma}$=1000} \\ \cline{3-5}
$00^+$ &$\chi^{\sigma 1}_{0}\chi^{f}_{d0}\chi^{c}_{d2}$,$\chi^{\sigma 2}_{0}\chi^{f}_{d0}\chi^{c}_{d1}$   &18457.6 &18337.8 &...  \\
$01^+$ &$\chi^{\sigma 5}_{1}\chi^{f}_{d0}\chi^{c}_{d1}$    &18526.6 &18650.4 &$-123.8$ \\
$02^+$ &$\chi^{\sigma 6}_{2}\chi^{f}_{d0}\chi^{c}_{d1}$  &18551.2 &18650.4 &$-99.2$ \\  \hline
 &  &  \multicolumn{3}{c}{$m_{\sigma}$=1500} \\ \cline{3-5}
$00^+$ &$\chi^{\sigma 1}_{0}\chi^{f}_{d0}\chi^{c}_{d2}$,$\chi^{\sigma 2}_{0}\chi^{f}_{d0}\chi^{c}_{d1}$  &18488.3   &18321.0 &... \\
$01^+$ &$\chi^{\sigma 5}_{1}\chi^{f}_{d0}\chi^{c}_{d1}$ &18566.3 &18654.0 &$-87.7$ \\
$02^+$ &$\chi^{\sigma 6}_{2}\chi^{f}_{d0}\chi^{c}_{d1}$ &18595.3 &18654.0 &$-58.7$ \\ \hline
 &  &  \multicolumn{3}{c}{$m_{\sigma}$=2500} \\ \cline{3-5}
$00^+$ &$\chi^{\sigma 1}_{0}\chi^{f}_{d0}\chi^{c}_{d2}$,$\chi^{\sigma 2}_{0}\chi^{f}_{d0}\chi^{c}_{d1}$   &18626.9 &18335.2 &...  \\
$01^+$ &$\chi^{\sigma 5}_{1}\chi^{f}_{d0}\chi^{c}_{d1}$     &18712.5 &18692.8 &... \\
$02^+$ &$\chi^{\sigma 6}_{2}\chi^{f}_{d0}\chi^{c}_{d1}$  &18745.3
&18692.8 &... \\  \hline
 &  &  \multicolumn{3}{c}{$m_{\sigma}$=3500} \\ \cline{3-5}
$00^+$ &$\chi^{\sigma 1}_{0}\chi^{f}_{d0}\chi^{c}_{d2}$,$\chi^{\sigma 2}_{0}\chi^{f}_{d0}\chi^{c}_{d1}$ &18767.6 &18371.8 &... \\
$01^+$ &$\chi^{\sigma 5}_{1}\chi^{f}_{d0}\chi^{c}_{d1}$&18849.4 &18736.0 &...\\
$02^+$ &$\chi^{\sigma 6}_{2}\chi^{f}_{d0}\chi^{c}_{d1}$&18880.7 &18736.0 &... \\
\hline \hline
\end{tabular}
\end{center}
\end{table}


\section{Summary}
\label{epilogue}

In the chiral quark model, we calculated the eigenenergies of the
low-lying $bb\bar{b}\bar{b}$ states with quantum numbers
$IJ^P=00^+, 01^+, 02^+$ using the Gaussian expansion method. Two
structures: meson-meson and diquark-antidiquark, and the coupling
of them are investigated. For the beauty-full system, the
interaction from the exchange of Goldstone bosons is absent
generally, and we found that the energies of $bb\bar{b}\bar{b}$
states with both structures are all higher than the corresponding
thresholds, leaving no space for a bound state in this situation.
Gluon exchange, as a short range force, it helps to form compact
tetraquarks rather than meson-meson molecules if bound four-quark
states do exit. In our calculations, $bb\bar{b}\bar{b}$ tetraquark
state may be a resonance state with the lowest mass $18872.8$ MeV
when considering the coupling of the meson-meson and
diquark-antidiquark structures. It is expected that the exotic
tetraquark states composed of four heavy quarks may be observed at
LHC.

As a test, the heavy meson $\eta_b$ exchange between the $b$ quarks is introduced,
its effect is too small to change the situation. However, the employment of
an effective scalar $\sigma$ will produce bound states for $bb\bar{b}\bar{b}$ system,
because of the universal attractive property of the $\sigma$ meson exchange.
The binding energy is smaller and smaller with the increasing the mass of $\sigma$.
Hopefully, our study will be helpful to searching for the exotic
tetraquark states composed of four heavy quarks.

\acknowledgments

This work is supported partly by the National Science Foundation
of China under Contract No. 11847145.



\end{document}